\PassOptionsToPackage{dvipsnames}{xcolor}
\documentclass[10pt,conference]{IEEEtran} 
\usepackage{balance}

\newcommand*{\datasize}{400\xspace}
\newcommand*{\sizeperrepo}{100\xspace}
\newcommand*{\numofwarnings}{39\xspace}
\newcommand*{\numofwarningspone}{18\xspace}
\newcommand*{\numoftp}{24\xspace}
\newcommand*{\numoftppone}{12\xspace}
\newcommand*{\numoffp}{15\xspace}
\newcommand*{\precision}{0.62\xspace}
\newcommand*{\precisionpone}{0.67\xspace}
\newcommand*{\mutationscore}{0.70\xspace}
\newcommand*{\mutationscorepone}{0.64\xspace}
\newcommand*{\numofcodebug}{17\xspace}
\newcommand*{\numofdocbug}{7\xspace}

\newcommand*{\numofunkbug}{12\xspace}
\newcommand*{\numofknownbug}{5\xspace}

\newcommand*{\numoffix}{11\xspace}

\newcommand*{\numofmorebug}{17\xspace}

\usepackage{xcolor}
\usepackage{url}
\usepackage{amsmath,amsfonts,amssymb}
\usepackage{rotating}
\usepackage{graphicx}
\usepackage{textcomp}
\usepackage{mathtools}
\usepackage{diagbox}
\usepackage{multirow}
\usepackage{pbox}
\usepackage{algorithm}
\usepackage{algorithmic}

\usepackage{listings}
\usepackage{enumerate}
\usepackage[shortlabels]{enumitem}
\usepackage[normalem]{ulem}
\usepackage{caption}
\usepackage{tcolorbox}
\usepackage{subcaption}
\usepackage{framed}
\usepackage{xspace}
\usepackage{booktabs}
\usepackage{cite}
\usepackage[hidelinks]{hyperref}
\usepackage{siunitx}

\newcommand*{\patchguru}{PatchGuru\xspace}
\newcommand*{\testora}{Testora\xspace}
\newcommand*{\rqone}{RQ$_1$\xspace}
\newcommand*{\rqtwo}{RQ$_2$\xspace}
\newcommand*{\rqthree}{RQ$_3$\xspace}
\newcommand*{\rqfour}{RQ$_4$\xspace}

\definecolor{deepblue}{rgb}{0,0,0.5}
\definecolor{deepred}{rgb}{0.6,0,0}
\definecolor{deepgreen}{rgb}{0,0.5,0}

\DeclareFixedFont{\ttb}{T1}{txtt}{bx}{n}{12} %
\DeclareFixedFont{\ttm}{T1}{txtt}{m}{n}{12}  %

\newcommand{\code}[1]{\texttt{\small #1}}

\begin{document}

\title{\patchguru: Patch Oracle Inference\\ from Natural Language Artifacts}

\author{%
\IEEEauthorblockN{Thanh Le-Cong$^{1}$, Bach Le$^{2}$, Toby Murray$^{2}$, Cristian Cadar$^{3}$, Michael Pradel$^{4}$\thanks{Contact: congthanh\_le@sutd.edu.sg, bach.le@unimelb.edu.au, toby.murray@unimelb.edu.au, c.cadar@imperial.ac.uk, michael@binaervarianz.de}}
\IEEEauthorblockA{$^{1}$Singapore University of Technology and Design (SUTD), Singapore}
\IEEEauthorblockA{$^{2}$University of Melbourne, Melbourne, Australia}
\IEEEauthorblockA{$^{3}$Imperial College London, London, United Kingdom}
\IEEEauthorblockA{$^{4}$CISPA Helmholtz Center for Information Security, Saarbr\"{u}cken, Germany}
}

\maketitle

\begin{abstract}
As software systems evolve, code changes may inadvertently introduce unintended behavioral changes.
Validating patches against their intended behaviors remains challenging due to the lack of adequate and machine-checkable specifications: regression tests are often incomplete, while natural language (NL) descriptions of patch intent are informal and non-executable.
This paper presents \patchguru, the first automated technique that infers executable patch specifications from real-world pull requests (PRs).
Given a PR, \patchguru leverages large language models to distill developer intent from its NL artifacts and synthesizes an under-approximate yet practical form of patch specifications called \emph{patch oracles}.
The benefits of patch oracles are threefold:
(1)~they concentrate on behaviors affected by the patch, making their inference more tractable and their validation more efficient;
(2)~they are expressed as runtime assertions, enabling automated validation; and
(3)~they are written within comparison programs combining both pre- and post-patch versions, thus allowing cross-version properties to be specified.
By comparing the behavior of pre- and post-patch versions of modified functions against these inferred oracles, \patchguru iteratively refines the oracles, identifies inconsistencies when violations arise, filters them via a self-review mechanism, and eventually reports likely bugs to developers.
We evaluate \patchguru on \datasize recent PRs from four widely used open-source Python projects. The evaluation shows that PatchGuru reports \numofwarnings warnings with a precision of \precision, yielding \numoftp confirmed true positives, including \numofunkbug previously unknown bugs, \numoffix of which have already been fixed following our reports.
Compared to the state-of-the-art patch validation technique \testora, \patchguru detects \numofmorebug more bugs (24 vs.\ 7) while increasing precision from 0.32 to \precision.
\patchguru also complements state-of-the-art LLM-based code review tools, detecting 12 bugs that Codex misses, while achieving substantially higher precision.
\patchguru has a reasonable average cost of around 8.9 minutes and \$0.07 per PR.
We envision \patchguru as a complement to existing code review and regression testing practices, providing both explicit documentation and automated validation of patch intent.
\end{abstract}

\section{Introduction}
\label{sec:intro}

Software systems continuously evolve to meet the changing needs of users and businesses.
This evolution is primarily realized through patches that aim to fix bugs, introduce new functionality, improve performance, or refactor code.
Ironically, patches, which are intended to improve the software, can inadvertently introduce  bugs and even vulnerabilities, leading to catastrophic consequences.
For example, a minor update in CrowdStrike caused a severe regression that disrupted approximately 8.5 million devices~\cite{CrowdStrikeOutage2024Wiki}, and a small patch in OpenSSL caused havoc by introducing the HeartBleed security vulnerability~\cite{heartbleed}.
Therefore, validating patches against their intended behavior before integration is crucial.

Patch validation faces a long-standing challenge: the lack of a precise, machine-checkable specification of intended patch behavior.
Regression tests are a common partial substitute, but they are costly to maintain and often incomplete in practice~\cite{covrig,covrig2,icse2026-ChaCo}.
Because they are largely not designed for individual patches, they may miss whether intended behavioral changes are actually realized.
This ``specification gap'' lets buggy patches pass existing tests while introducing unintended behavior or preserving behavior meant to change.

A common proxy for intent is natural language (NL) artifacts, such as commit messages and pull request (PR) descriptions.
Yet, these artifacts are informal and not directly machine-checkable for automated patch validation.
As a result, developers rely on manual code review~\cite{Bacchelli2013ExpectationsOA, McIntosh2015AnES}, which is time-consuming, error-prone, and dependent on reviewer expertise.
Moreover, most review comments focus on maintainability rather than functional correctness~\cite{Bacchelli2013ExpectationsOA}, allowing intent-misaligned buggy patches to slip into production and potentially cause regressions and vulnerabilities~\cite{Khoshnoud2022WhichBA}.
More recently, LLM-based code review tools have been adopted in practice~\cite{cihan2025automated}.
However, similar to human reviewers, these tools primarily reason about code patches implicitly rather than explicitly validating their runtime behavior, and therefore cannot reliably determine whether a patch's actual behavior matches its stated intent~\cite{lin2025codereviewqa}.
Across these approaches, a fundamental limitation remains: they do not produce an explicit, machine-checkable specification of the behavioral change that a patch is intended to introduce. 
Such a specification is crucial for automatically validating intended patch behavior against actual runtime behavior, rather than inferring it indirectly from informal NL descriptions or reviewer judgments.

Recently, Testora~\cite{Pradel2025TestoraUN} directly targets this specification gap, generating differential tests to expose behavioral differences between the pre- and post-patch versions and using an LLM to determine whether the differences are intended.
However, as Testora relies on a differential oracle, it cannot identify bugs where a patch fails to implement an intended change and both versions exhibit the same behavior. 
More importantly, developer intent is represented implicitly through LLM judgments rather than as an explicit specification that can be validated, refined, or shared with developers. 
Thus, although Testora goes beyond static code review by observing runtime behaviors, it still does not provide a machine-checkable specification of what the patch is supposed to change.

The need for the machine-checkable specification in patch validation has been articulated in a recent vision paper by Cadar et al.~\cite{Cadar2023PatchSV}, which proposed \emph{patch specifications} to formally specify the intended behavioral delta between pre- and post-patch program versions.
However, patch specifications are designed to hold universally over all inputs. 
This makes them highly desirable, but difficult to construct in practice, as they require precise reasoning about the intended behavior of a code change across relevant inputs and edge cases. 
Consequently, patch specifications are typically written manually~\cite{Cadar2023PatchSV,Sharma2025P3Reasoning}, and no existing technique automatically infers them from real-world patches, limiting their practical adoption.

This paper presents \patchguru, the first practical, automated technique for generating machine-checkable patch specifications from real-world patches, and checking them via automatically generated \emph{comparison programs}~\cite{Lars2025ChangeGuard}.
\patchguru infers what we call \emph{patch oracles}, an under-approximate yet practical form of patch specifications that capture key behavioral properties of patches.
We design patch oracles as \textit{runtime assertions} paired with \textit{concrete inputs} within a comparison program, which is a synthetic program that contains the pre- and post-patch versions of a modified function as independent functions, allowing for direct comparison of their behavior. 
The benefits of this design are threefold.
First, patch oracles focus on behavior affected by the patch, making their inference more tractable and their validation more efficient compared to whole-program validation, e.g., regression testing or formal verification.
Second, their representation as assertions enables direct execution within comparison programs, providing automated patch validation through dynamic analysis.
Third, the use of comparison programs allows \patchguru to specify cross-version properties between pre- and post-patch behaviors, which are essential for characterizing patch intent, yet difficult to express using traditional specification approaches.

\smallskip\noindent\textbf{Example.} Consider a buggy patch that aims to change a function \texttt{f} from the identity function to one that returns its argument plus 1.
\patchguru automatically generates the following \emph{comparison program}, embedding both versions as independent functions, together with a \emph{patch oracle} consisting of runtime assertions paired with test inputs to be checked.
\begin{lstlisting}[numbers=none, xleftmargin=0.1cm, basicstyle=\ttfamily\scriptsize, aboveskip=2pt, belowskip=2pt]
def pre_f(x): return x         # pre-patch version
def post_f(x): return x + 2    # post-patch version
test_inputs = [1, 2, 3]    
for x in test_inputs:
    assert post_f(x) == pre_f(x) + 1 
\end{lstlisting}
Executed over multiple inputs $x$, a violation of the oracle immediately exposes the inconsistency, i.e.,\ the fact that the patched function mistakenly returns \code{x + 2} instead of \code{x + 1}.
\S\ref{sec:approach} and Fig.~\ref{fig:example} present a detailed real-world example.

To infer patch oracles, we utilize NL artifacts associated with patches, e.g., commit messages, descriptions, and discussions, which typically convey the developers' intent behind a change.
Given a PR, \patchguru distills relevant NL artifacts and leverages LLMs to synthesize a machine-checkable patch oracle and an associated comparison program.
The inferred patch oracle is executed and iteratively refined by generalizing assertions and exploring edge cases. 
When assertion failures arise, \patchguru adopts an LLM-as-a-judge to review the failures, confirm inconsistencies between the code changes and inferred oracles, and generate a report for developers.

Our evaluation applies \patchguru to \datasize real-world PRs from four popular and complex open-source projects, showing that it finds bugs that have remained undetected during code review and regression testing.
In total, \patchguru identifies \numoftp bugs, i.e., true inconsistencies between code changes and their intended behavior, split between \numofcodebug code bugs and \numofdocbug documentation bugs.
Among the code bugs, \numofunkbug were previously unknown, all have been confirmed by developers, and \numoffix are already fixed following our reports.
Compared to Testora~\cite{Pradel2025TestoraUN}, \patchguru detects \numofmorebug more bugs (24 vs.\ 7) while reducing the false positive rate from 0.68 to 0.38.
Compared to the code review feature of Codex CLI~\cite{openai_codex_review}, a commercial LLM-based coding agent from OpenAI, \patchguru achieves substantially higher precision and detects 12 bugs that Codex misses, showing that \patchguru can complement existing LLM-based code review tools to improve bug detection and provide actionable warnings to developers.
Additionally, in a fault injection campaign, \patchguru-inferred oracles achieve a higher mutation score than existing regression tests (0.70 vs.\ 0.58), while combining the two achieves a mutation score of 0.81, demonstrating that \patchguru-inferred oracles effectively complement existing regression tests.
Finally, we show that the costs of using \patchguru are acceptable for real-world deployment, with an average inference time of 8.9 minutes and an average LLM cost of USD 0.07 per PR.

In summary, this paper makes the following contributions:
\begin{itemize}
    \item \textit{Idea.} We introduce the idea of patch oracle inference from NL artifacts associated with patches, providing a novel way of bridging the specification gap in patch validation.
    \item \textit{Technique.} We present \patchguru, an automated technique that synthesizes and iteratively refines patch oracles as comparison programs. %
    \item \textit{Evaluation.} We conduct a large-scale evaluation of \patchguru on \datasize real-world PRs from popular open-source Python projects, demonstrating its effectiveness in inferring adequate patch oracles and identifying real bugs. %
    \item \textit{Artifacts.} We open-source our code and data~\cite{Anom2025PatchGuruReplicationPackage}.
\end{itemize}

\section{Approach}
\label{sec:approach}

\subsection{Problem Statement}
\label{sec:problem}

Before presenting our approach, we formally define the problem that \patchguru aims to solve.

\textbf{Input:} \patchguru takes as input a PR consisting of two components:
(1)~code changes $\mathcal{C}$, and
(2)~NL artifacts $\mathcal{D}$, including the PR description, commit messages, and related discussions that explain the rationale for the change. 
We currently focus on PRs that modify exactly one function, which simplifies the construction of comparison programs. We acknowledge this as a limitation and further discuss it in \S\ref{sec:discussion}.

\textbf{Output:} \patchguru outputs a triplet $(\mathcal{O}, \mathcal{P}, \mathcal{R})$, where
~$\mathcal{O}$ is an inferred patch oracle comprising runtime assertions and test inputs that specify the intended relationship between pre- and post-patch behavior; 
~$\mathcal{P}$ is a comparison program that validates both versions against $\mathcal{O}$; and 
~$\mathcal{R}$ is a natural language report explaining detected inconsistencies. 
When an inconsistency is found, \patchguru raises a warning and returns the triplet as reproducible evidence. 
We assume the pre-patch version is correct, as the goal is to detect patch-introduced bugs.

\subsection{Overview}

\begin{figure}
    \centering
    \includegraphics[width=\columnwidth]{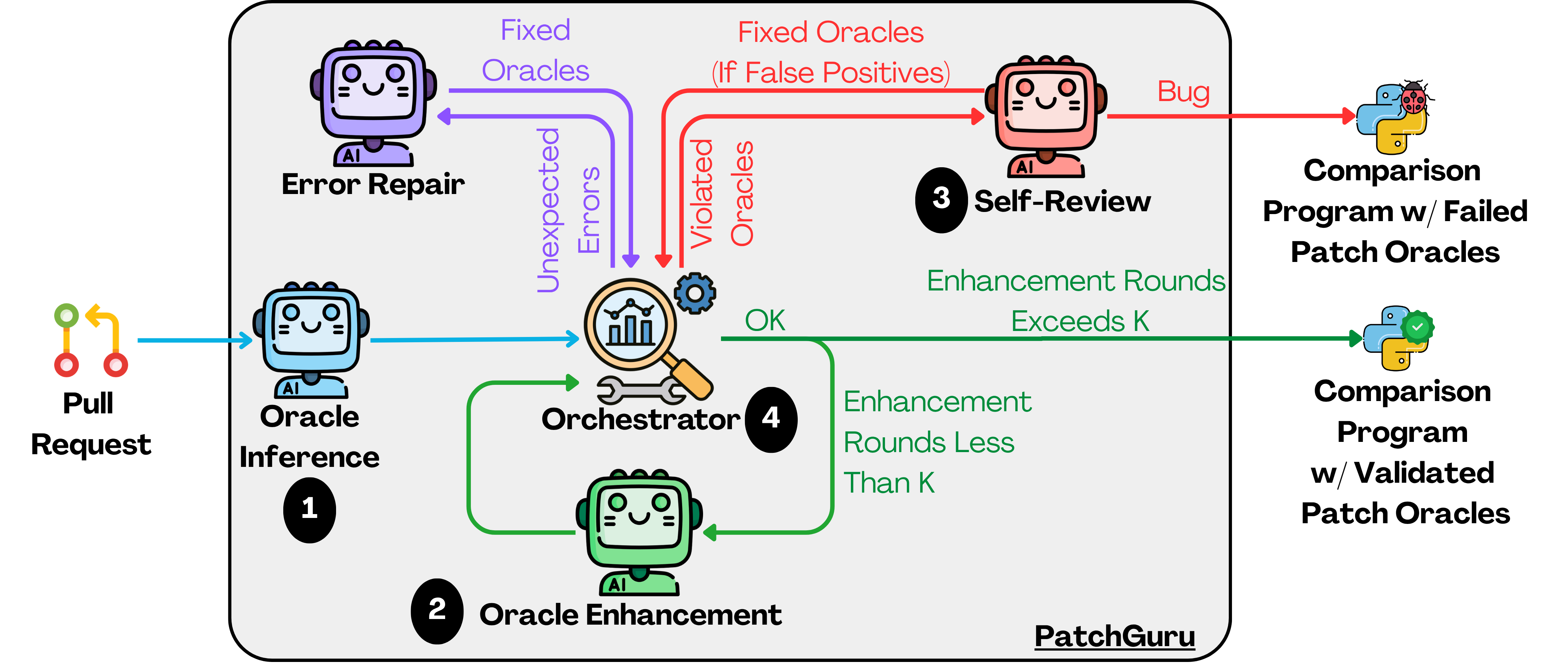}
    \caption{Overview of \patchguru's workflow.}
    ~\label{fig:approach}
\end{figure}

Fig.~\ref{fig:approach} presents an overview of \patchguru's workflow, which consists of three main phases coordinated by an orchestrator.
Given a PR, \patchguru first employs an LLM-based inference module (\textcircled{1}, \S\ref{sec:approach_inference}) to derive an initial patch oracle from the PR's NL artifacts.
The inferred oracle includes runtime assertions that encode the intended behavior described in the PR, together with test inputs that trigger these behaviors. 
If the behavior of the code patch is consistent with the inferred oracle, \patchguru enters an iterative oracle enhancement phase (\textcircled{2}, \S\ref{sec:approach_enhancement}) to improve the completeness of the inferred oracle.
During this phase, \patchguru generalizes initial assertions to cover broader input domains, and synthesizes additional test inputs to explore edge cases.
At any point, if inconsistencies are detected, \patchguru triggers a self-review phase (\textcircled{3}, \S\ref{sec:approach_review}), which applies an LLM-as-a-judge mechanism to distinguish true inconsistencies from false positives.
Throughout the workflow, the orchestrator (\textcircled{4}, \S\ref{sec:approach_analyzer}) executes inferred oracles, monitors runtime behavior, repairs incorrect oracles, and orchestrates transitions between phases.

\begin{figure}[t]
\centering
\begin{framed}
\footnotesize
\raggedright
\textbf{(a) PR Context}

{\scriptsize
\underline{PR Title}: Fix: add file URL handling to the URL validator

\vspace{0.2em}

\underline{PR Description:} Requires a modicum of special handling due to hostnames being optional.
Fixes \#2249. [Issue \#2249] Marshmallow's \texttt{URL} validator does not accept file URLs without host. For example, \texttt{file:///var/storage/somefile.zip} raises a \texttt{ValidationError}.
}

\vspace{0.3em}

\textbf{(b) Inferred Patch Oracle \& Comparison Program}
\begin{lstlisting}[backgroundcolor=\color{blue!10},label={lst:validation},aboveskip=3pt,belowskip=0pt]
from marshmallow.fields import URL
from regex import search
from marshmallow.exceptions import ValidationError
def pre_fn(validator, url):
    # pre-patch implementation of fields.URL.__call__
def post_fn(validator, url):
    # post-patch implementation of fields.URL.__call__
def call(validator, pre_fn, post_fn, url):
    try:
        pre_res, pre_exc = pre_fn(validator, url), None
    except ValidationError as e:
        pre_res, pre_exc = None, e
    try:
        post_res, post_exc = post_fn(validator, url), None
    except ValidationError as e:
        post_res, post_exc = None, e
    return pre_res, pre_exc, post_res, post_exc
\end{lstlisting}%
\begin{lstlisting}[backgroundcolor=\color{green!10},nolol,firstnumber=last,aboveskip=0pt,belowskip=0pt]
validator = URL(schemes=["https", "file"])(*@\label{lst:validator-init}@*)
## PRESERVED: VALID URLS
valid_url = "http://example.com/path"
pre_res, pre_exc, post_res, post_exc = call(validator, pre_fn, post_fn, valid_url)
assert pre_exc is None and post_exc is None and pre_res == valid_url and post_res == valid_url, ("[PRESERVED] Both pre_fn and post_fn should accept valid URLs.")
## PRESERVED: INVALID URLS
invalid_url = "ftp://example.com/path"
pre_res, pre_exc, post_res, post_exc = call(validator, pre_fn, post_fn, invalid_url)
assert isinstance(pre_exc, ValidationError) and isinstance(post_exc, ValidationError), ("[PRESERVED] Both pre_fn and post_fn should reject invalid URLs.")
## CHANGED: FILE URLS
lower_file = "file:///etc/passwd"
pre_res, pre_exc, post_res, post_exc = call(validator, pre_fn, post_fn, lower_file)
assert isinstance(pre_exc, ValidationError) and post_exc is None and post_res == lower_file, ("[CHANGED] post_fn should accept file URLs while pre_fn rejects them.")(*@\label{lst:oracle-lower}@*)
\end{lstlisting}%
\begin{lstlisting}[backgroundcolor=\color{orange!15},nolol,firstnumber=last,aboveskip=0pt,belowskip=2pt]
## CHANGED: FILE URLS (ENHANCED)
upper_file = "FILE:///etc/passwd"
pre_res, pre_exc, post_res, post_exc = call(validator, pre_fn, post_fn, upper_file)
assert isinstance(pre_exc, ValidationError) and post_exc is None and post_res == upper_file, ("[CHANGED] post_fn should accept file URLs (case-sensitive) while pre_fn rejects them.")(*@\label{lst:oracle-upper}@*)
\end{lstlisting}

\vspace{0.2em}

\underline{Error Message:} \texttt{\scriptsize AssertionError: [CHANGED] post\_fn should accept file URLs while pre\_fn rejects them.}

\vspace{0.2em}

\underline{Self-Review:} The failing assertion is caused by the post-patch code rejecting uppercase \texttt{FILE:///...} URLs even though the PR intent is to support file URLs without hostnames. Conclusion: [BUG]
\end{framed}
\caption{Shortened real-world example based on a Marshmallow patch for function \code{fields.URL}.
  It shows (a)~PR context and (b)~comparison program ({\setlength{\fboxsep}{1.5pt}\colorbox{blue!10}{\scriptsize blue}}) and patch oracle ({\setlength{\fboxsep}{1.5pt}\colorbox{green!10}{\scriptsize green}}: initial inference; {\setlength{\fboxsep}{1.5pt}\colorbox{orange!15}{\scriptsize orange}}: after enhancement) inferred by \patchguru. The PR introduces a bug detected by \patchguru.
}
\label{fig:example}
\end{figure}

\begin{table*}[t]
\centering
\small
\caption{\patchguru's prompt design including tasks, key guidelines and inputs. See full prompts in \cite{online_appendix}. 
}
\label{tab:prompt-summary}
\setlength{\tabcolsep}{3pt}
\setlength{\aboverulesep}{1pt}
\setlength{\belowrulesep}{1pt}
\resizebox{\textwidth}{!}{
\begin{tabular}{@{}
>{\raggedright\arraybackslash}p{1.90cm}
>{\raggedright\arraybackslash}p{4cm}
>{\raggedright\arraybackslash}p{10cm}
>{\centering\arraybackslash}p{0.45cm}
>{\centering\arraybackslash}p{0.45cm}
>{\centering\arraybackslash}p{0.45cm}
>{\centering\arraybackslash}p{0.45cm}
>{\centering\arraybackslash}p{0.45cm}
>{\centering\arraybackslash}p{0.45cm}@{}}
\toprule
\textbf{Module} & \textbf{Task} & \textbf{Key guidelines} & \multicolumn{6}{c}{\textbf{Inputs}} \\
\cmidrule(lr){4-9}
& & & \textbf{NL}\textsuperscript{a} & \textbf{ctx}\textsuperscript{b} & \textbf{chg}\textsuperscript{c} & \textbf{ora}\textsuperscript{d} & \textbf{pro}\textsuperscript{e} & \textbf{rpt}\textsuperscript{f} \\
\midrule
Inference & Infer initial patch oracle and comparison program & Understand PR context and developer intent; derive invariant behaviors and intended changes; produce precise, executable assertions and comparison program with test inputs & \checkmark & \checkmark & & & & \\
Enhancement & Generalize and broaden coverage of patch oracle & Generalize existing assertions; explore edge cases and boundary conditions; diversify inputs while keeping existing assertions intact & \checkmark & \checkmark & & \checkmark & \checkmark & \\
Self-review & Classify assertion errors as bugs or false positives  & Cross-check PR details, code changes, and failure evidence; distinguish true bugs from flawed tests; provide a clear conclusion and correction if needed & \checkmark & \checkmark & \checkmark & \checkmark & \checkmark & \checkmark \\
Repair & Fix runtime errors & Inspect the full traceback; check common errors; repair observed issues & \checkmark & \checkmark & & \checkmark & \checkmark & \checkmark \\
\bottomrule
\multicolumn{9}{l}{\footnotesize \textsuperscript{a}NL artifacts~(\S\ref{sec:approach_inference_nl});\enspace \textsuperscript{b}code context~(\S\ref{sec:approach_inference_code});\enspace \textsuperscript{c}chg\,=\,code changes;\enspace \textsuperscript{d}ora\,=\,current patch oracle;\enspace \textsuperscript{e}pro\,=\,comparison program;\enspace \textsuperscript{f}execution report~(\S\ref{sec:approach_analyzer_execution}).} \\
\end{tabular}
}
\end{table*}

\subsection{Oracle Inference}
\label{sec:approach_inference}

The goal of the oracle inference phase is to formalize the developer's intent, as expressed in the NL artifacts of a PR, into a machine-checkable patch oracle.
The resulting oracle consists of runtime assertions paired with test inputs that characterize the intended behavior of the modified function.
Specifically, we define three general assertion categories to capture the behavioral relationship between the pre-patch and post-patch versions of the function as follows:
(1)~\emph{preserved behavior} assertions that specify identical behavior between the pre- and post-patch versions for certain inputs;
(2)~\emph{changed behavior} assertions that specify modified behaviors for specific inputs; and
(3)~\emph{new behavior} assertions that describe entirely new functionality added by the patch.
Given a PR, we prompt an LLM to synthesize concrete assertions and corresponding test inputs through a three-step process:
(1)~collecting relevant NL artifacts, 
(2)~extracting the code context, and 
(3)~inferring patch oracles from the collected information.

\subsubsection{Gathering Natural Language Artifacts}
\label{sec:approach_inference_nl}

\patchguru gathers NL artifacts associated with the PR, including the PR title, description, and developers' discussions 
(see Fig.~\ref{fig:example}(a) for an example) as they often contain explicit descriptions of the intended behavior and rationale behind the code changes.
It also retrieves issues linked in the PR, as they may contain intent not explicitly stated in the PR description.
For instance, in Fig.~\ref{fig:example}(a), the original PR description does not specify that \texttt{fields.URL} should accept file URLs without hostnames; this intent was instead documented in the referenced issue \#2249.
To do so, \patchguru extracts issue URLs from the PR description via regular expression matching and retrieves the corresponding issue content via GitHub's REST API.
As these references may contain irrelevant information, \patchguru applies an LLM-based distillation to retain only details that directly relate to the intended changes, discarding unrelated information and tangential discussions.

\subsubsection{Extracting Context of Modified Functions}
\label{sec:approach_inference_code}

\patchguru also extracts relevant code context for the changed function from the codebase
to provide additional information for oracle inference and ensure the executability of the inferred oracle.
In particular, \patchguru first identifies the modified function by analyzing the code changes in the PR, and then
identifies dependencies (i.e., callees) of the modified function, which are necessary for executing the inferred patch oracle. %
For dependencies within the same file, \patchguru parses the modified file into an abstract syntax tree and extracts top-level function and class definitions.
For dependencies in external modules, \patchguru analyzes import statements and retrieves the corresponding modules.
If the modified function is a class method, \patchguru also extracts the complete class definition to provide additional context and ensure the correct initialization of instance variables during oracle execution.
For example, in Fig.~\ref{fig:example}(b), \patchguru extracts the complete implementation of the \texttt{fields.URL.\_\_call\_\_} method, which is the modified function, along with its dependencies such as the \texttt{search} function from the \texttt{regex} library, the \texttt{ValidationError} class from the \texttt{marshmallow.exceptions} module, and the class definition of \texttt{URL} to provide necessary context for initializing the \texttt{URL} validator instance in line~\ref{lst:validator-init}.

\subsubsection{Patch Oracle Inference}
\label{sec:approach_inference_main}

Using the collected NL artifacts and code context, \patchguru invokes an LLM with a structured prompt template, consisting of a task description, guidelines, and contextual inputs 
(summary in Table~\ref{tab:prompt-summary}; full template in~\cite{online_appendix}), to synthesize an initial patch oracle.
A key design choice in our prompt is to provide step-by-step guidelines, inspired by the least-to-most prompting technique~\cite{zhou2023leasttomost}, decomposing the patch-oracle inference task into manageable subtasks to improve the quality and reliability of the generated oracles.
In addition to the patch oracle, we also ask the LLM to jointly generate a comparison program to enable automatic validation of the generated oracle. 
Specifically, \patchguru instructs it to generate the comparison program with placeholders for the pre- and post-PR implementations, which are later replaced with the actual implementations extracted from the codebase as described in \S\ref{sec:approach_inference_code} to ensure executability.
For example, Fig.~\ref{fig:example}(b) shows the comparison program and patch oracle inferred by \patchguru for the PR in Fig.~\ref{fig:example}(a).
The oracle asserts three behavioral properties:
(1)~valid URLs are accepted unchanged by both versions and returned as-is;
(2)~invalid URLs raise a \texttt{ValidationError} in both versions; and
(3)~file URLs without hostnames are rejected pre-patch but accepted post-patch.

\subsection{Oracle Enhancement}
\label{sec:approach_enhancement} 

Given the initial patch oracle, the oracle enhancement phase iteratively refines it to improve its bug detection capabilities. 
For instance, in Fig.~\ref{fig:example}, \patchguru initially infers assertions that account only for lowercase file URLs, i.e., those beginning with ``file://'' (line~\ref{lst:oracle-lower}).
During the oracle enhancement phase, \patchguru generalizes these assertions to include uppercase file URLs, e.g., those beginning with ``FILE://'' (line~\ref{lst:oracle-upper}), improving oracle adequacy and exposing a previously unknown PR-introduced bug: the post-patch code fails to handle uppercase file URLs correctly.

To enhance the inferred oracle, \patchguru invokes an LLM with a prompt that has a similar structure to the oracle inference prompt, but with a different task and guidelines (summary in Table~\ref{tab:prompt-summary}; full template in~\cite{online_appendix}).
The new task and guidelines are designed to address the primary failure modes of initial oracles as identified in our preliminary analysis: 
they tend to be overly specific to the inferred example inputs (mitigated by generalizing existing assertions and diversifying test inputs) 
and may fail to capture boundary behaviors (mitigated by exploring edge cases and boundary conditions).

\subsection{Self-Review}\label{sec:approach_review}

Once assertions are violated during execution, \patchguru enters the self-review phase to determine whether these violations correspond to 
true inconsistencies (i.e., bugs) or false positives arising from incomplete or incorrect oracles.
Specifically, for each failing assertion, \patchguru invokes an LLM-as-a-judge with a self-review prompt 
(summary in Table~\ref{tab:prompt-summary}; full template in~\cite{online_appendix})
that instructs it to analyze the NL artifacts, code context, code changes, and execution report to provide a verdict of either true positive or false positive, along with a detailed explanation supporting the verdict.
In case of a false positive, \patchguru also prompts the LLM to propose concrete enhancements to the oracle. 
In case of a true positive, \patchguru raises a warning to notify developers of the detected inconsistency and generates a validation report that summarizes the findings.
For instance, in Fig.~\ref{fig:example}, the LLM identifies that the assertion failure arises from the post-patch code's case-sensitive handling of file URLs, which contradicts the developer's intent.
Thus, the LLM identifies the failure as a true inconsistency, i.e., a bug, and provides a detailed explanation.

\begin{algorithm}[t]
\footnotesize
\caption{Core algorithm of \patchguru.
}\label{alg:core}
\begin{algorithmic}[1]
  \REQUIRE PR $(\mathcal{C}, \mathcal{D})$ with code patch $\mathcal{C}$ and NL artifacts $\mathcal{D}$,\\
  maximum number of LLM calls $M$,
  maximum number of enhancement iterations $N$
\ENSURE Patch oracle $\mathcal{O}$, comparison program $\mathcal{P}$, and validation report $\mathcal{R}$

\STATE $q \leftarrow 0$ \COMMENT{Track LLM calls}

\STATE $\mathcal{O}, q\leftarrow \mathit{INFER}(\mathcal{D})$ \COMMENT{Infer initial patch oracle (\S\ref{sec:approach_inference})}
\STATE $\mathit{status}, \mathit{msg} \leftarrow \mathit{EXECUTE}(\mathcal{O}, \mathcal{C})$ \COMMENT{Execute patch oracle (\S\ref{sec:approach_analyzer_execution})}

\STATE $\mathit{iter} \leftarrow 0$
\WHILE{$\mathit{iter} < N$}
    \IF{$\mathit{status}$ is \texttt{NO\_VIOLATION}}
        \STATE $\mathcal{O}, q_{u} \leftarrow \mathit{ENHANCE}(\mathcal{O}, \mathcal{D})$ \label{alg:line:enhance} \COMMENT{Enhance oracle (\S\ref{sec:approach_enhancement})}
        \STATE $\mathit{iter} \leftarrow \mathit{iter} + 1$
    \ELSIF{$\mathit{status}$ is \texttt{ASSERTION\_VIOLATION}}
        \STATE $\mathcal{O}, \mathcal{R}, \mathit{res}, q_{\mathit{u}}\leftarrow \mathit{REVIEW}(\mathcal{O}, \mathit{msg}, \mathcal{D}, \mathcal{C})$ \label{alg:line:review} \COMMENT{Self-review (\S\ref{sec:approach_review})}
        \IF{$\mathit{res}$ is \texttt{TRUE\_POSITIVE}} 
            \RETURN $\mathcal{O}, \mathcal{R}$
        \ENDIF
    \ELSIF{$\mathit{status}$ is \texttt{UNEXPECTED\_ERROR}}
        \STATE $\mathcal{O}, q_{u} \leftarrow \mathit{REPAIR}(\mathcal{O}, \mathit{msg}, \mathcal{D})$ \label{alg:line:repair} \COMMENT{Repair errors (\S\ref{sec:approach_repair})}
    \ENDIF
    \STATE $q \leftarrow q + q_{u}$ \COMMENT{update LLM call count}
    \IF{$q \geq M$}
        \RETURN $\mathcal{O}, \varnothing$ \COMMENT{Exceeded LLM call limit}
    \ENDIF
    \STATE $\mathit{status}, \mathit{msg} \leftarrow \mathit{EXECUTE}(\mathcal{O}, \mathcal{C})$ \COMMENT{Execute patch oracle (\S\ref{sec:approach_analyzer_execution})}
\ENDWHILE
\RETURN $\mathcal{O}, \varnothing$ \COMMENT{No inconsistencies found}
\end{algorithmic}
\end{algorithm}

\subsection{Core Algorithm}\label{sec:approach_analyzer}

The orchestrator manages the \patchguru workflow by managing interactions among LLM modules and executing the inferred oracles. 
Given a set of oracles inferred or refined by the LLM, the orchestrator
(i)~constructs and executes a comparison program with the oracles and
(ii)~selects subsequent actions based on the execution outcome, including oracle enhancement, self-review, or error repair, following Alg.~\ref{alg:core}.

\subsubsection{Constructing and Executing Comparison Programs}
\label{sec:approach_analyzer_execution}

As mentioned in \S\ref{sec:approach_inference_main}, the patch oracle contains placeholders for the pre- and post-patch implementations. 
The orchestrator replaces these placeholders by extracting the relevant code functions from the pre- and post-patch versions of the modified file, based on the ranges of deleted and added lines. 
To avoid naming conflicts, the orchestrator renames the pre- and post-patch functions by appending the prefixes \texttt{pre\_} and \texttt{post\_}. 
Additionally, the orchestrator also integrates all necessary dependencies extracted in \S\ref{sec:approach_inference_code} into the comparison program to ensure executability.
By following this procedure, the orchestrator converts the inferred patch oracle into an executable comparison program to validate pre- and post-patch behavior.
While our extraction may miss some dependencies due to the complexity of real-world codebases, \patchguru handles them in the subsequent error repair phase (\S\ref{sec:approach_repair}). 
As isolated execution of arbitrary modified functions is challenging, \patchguru may generate mock data types or stubs to make comparison programs executable, similar to ChangeGuard~\cite{Lars2025ChangeGuard}. Such synthetic inputs can be unrealistic and may cause false positives, as discussed in \S\ref{sec:rq1}.

Once an executable comparison program is constructed, the orchestrator proceeds to execute it while monitoring runtime behavior (Alg.~\ref{alg:core}).
To this end, \patchguru builds a Docker container that encapsulates the necessary runtime environment, including language interpreters, libraries, and dependencies required for executing the comparison program. 
This containerized approach ensures consistency and isolation during execution, mitigating potential conflicts with the host environment. During execution, the orchestrator captures detailed execution logs, including standard output, error messages, and stack traces. 
It also monitors the evaluation of assertions defined in the patch oracle, recording any violations or exceptions that occur. 
After execution, the orchestrator compiles a comprehensive report summarizing the execution results, including the status of each assertion (passed or failed), any runtime errors encountered, and relevant execution logs. 
This report is then relayed back to the respective LLM modules for oracle enhancement, self-review, or error repair.

\subsubsection{Selecting Next Actions}
\label{sec:approach_analyzer_next}

After executing the comparison program, the orchestrator evaluates the results to determine the next action, considering several scenarios:

\begin{itemize}

\item \emph{No Errors or Violations:} If all assertions pass without runtime errors, the orchestrator considers the pre- and post-patch behaviors consistent under the current oracle. 
\patchguru then invokes the oracle enhancement phase (\S\ref{sec:approach_enhancement}, Alg.~\ref{alg:core}, line~\ref{alg:line:enhance}) to further refine the oracle.
If this phase reaches its iteration limit without detecting inconsistencies, \patchguru terminates and reports the comparison program with the validated oracle.

\item \emph{Assertion Violations:} If any assertions fail during execution, the orchestrator triggers the self-review phase (\S\ref{sec:approach_review}; Alg.~\ref{alg:core}, line~\ref{alg:line:review}).
The self-review process determines whether the violations represent true inconsistencies or false positives that require oracle enhancement. 
Note that, as \patchguru also generates assertions on the pre-patch function to capture its intended behavior, assertion violations can also be raised on the pre-patch function.
However, as we assume that the pre-patch function is correct (\S\ref{sec:problem}), any assertion violations on the pre-patch function are treated as \emph{reasoning errors} made by the LLM and handled in the error repair phase (\S\ref{sec:approach_repair}, Alg.~\ref{alg:core}, line~\ref{alg:line:repair}). This design enables \patchguru to avoid misinterpreting the behavior of the pre-patch function, thereby providing more accurate patch oracles.

\item \emph{Unexpected Errors:} If any other errors occur during execution, the orchestrator categorizes these as execution errors and invokes the error repair module (\S\ref{sec:approach_repair}, Alg.~\ref{alg:core}, line~\ref{alg:line:repair}) to address them. Specifically, we consider two main categories of execution errors: (1)~\emph{Syntax errors}, such as incorrect indentation, missing colons, or unmatched parentheses; (2)~\emph{Runtime errors} during comparison-program execution, such as \texttt{NameError} from undefined variables or functions, and \texttt{TypeError} from incompatible data-type operations.

\end{itemize}

\subsubsection{Error Repair}
\label{sec:approach_repair}

When execution errors are detected by the orchestrator, \patchguru invokes an LLM-based error repair module to diagnose and resolve these issues.
As discussed above, \patchguru focuses on three main categories of errors:
(i)~Reasoning Errors, which lead to assertion violations on the pre-patch function,
(ii)~Syntax Errors during compilation, and
(iii)~Runtime Errors during execution.
To address these errors, \patchguru prompts an LLM with a structured error repair prompt that guides the LLM to analyze the error message and the context of the generated code to identify the root cause and propose concrete fixes.

\section{Evaluation}
\label{sec:evaluation}

We answer the following research questions (RQs):

\begin{itemize}
    \item[\textbf{\rqone}:] Can patch oracles generated by \patchguru detect real-world bugs?
    \item[\textbf{\rqtwo}:] How adequate are the patch oracles generated by \patchguru?
    \item[\textbf{\rqthree}:] What is the per-PR cost of \patchguru?
    \item[\textbf{\rqfour}:] How do individual components of \patchguru contribute to its performance?
\end{itemize}

\subsection{Implementation Details and Experimental Setup}
\label{sec:experimental-setup}

While \patchguru is conceptually language-agnostic, we evaluate it by implementing a prototype for the Python ecosystem. 
We select Python due to its popularity and the availability of mature, large-scale open-source repositories. 
Moreover, LLMs have demonstrated strong capabilities in understanding and generating Python code~\cite{Chen2021}.
We employ GPT-5-mini~\cite{openai_introducing_gpt5} as the underlying LLM as our preliminary experiments showed that it offers a
reasonable trade-off between effectiveness and computational cost.
As OpenAI does not allow configuring the temperature for GPT-5-mini, we use the default temperature setting hard-coded by OpenAI.
We set per-call limits of 400,000 input tokens and 16,384 output tokens; in our evaluation, most runs are well below these bounds. %

All experiments are conducted on an Ubuntu 22.04 LTS server with a 32-core AMD EPYC 9474F processor and \SI{128}{\giga\byte} of RAM.
To maintain computational efficiency, each test execution, including compilation and execution within Docker (as described in \S\ref{sec:approach_analyzer_execution}), is limited to \SI{1}{\hour}. 
Each phase of \patchguru is individually capped at 5 LLM invocations per PR, with an overall budget of 20 LLM invocations per PR across all phases.
This limit was empirically determined by our preliminary experiments to balance effectiveness and cost.

\textbf{Dataset.} To evaluate \patchguru, we select four widely-used open-source Python projects: Keras (deep learning), Marshmallow (serialization), Pandas (data analysis), and SciPy (scientific computing).
This selection aligns with the benchmark dataset used by our primary baseline \testora~\cite{Pradel2025TestoraUN} while covering diverse application domains.
From each project, we collect merged PRs and apply three filters.
First, we exclude PRs that modify only documentation (files in documentation directories, code comments, README files).
Second, we exclude PRs whose changes are confined exclusively to test files.
Third, we retain only PRs that modify exactly one function.
From the qualifying PRs, we select the \sizeperrepo{} most recently merged PRs per project, yielding \datasize{} PRs in total.
The complete list of PR identifiers is available in the folder (\texttt{cache/pr\_ids}) of our replication package~\cite{Anom2025PatchGuruReplicationPackage}.

\textbf{Baselines.} To evaluate the effectiveness of our approach, we compare \patchguru against three baselines: \testora, LLM-based code review, and developer-written regression tests.

\paragraph{\testora (\rqone)} \testora~\cite{Pradel2025TestoraUN} is a state-of-the-art patch validation technique that also checks Python code patches against NL artifacts in PRs. 
It generates tests and executes them independently on pre- and post-patch program versions to obtain their outputs for comparison. 
When differences arise between the outputs, \testora uses an LLM to determine whether observed behavioral differences are intended w.r.t.\ the NL artifacts; if not, it raises a warning.
For a fair comparison, we sought the help of \testora's author to run it on our dataset of PRs using their recommended settings and GPT-5-mini as the underlying LLM (i.e., the same model used by \patchguru).
We compare against \testora in \rqone but not in \rqtwo because \testora generates test cases rather than explicit patch oracles, making it infeasible to compare their adequacy.

\paragraph{LLM-based code review (\rqone)}
We also compare \patchguru against OpenAI's Codex CLI~\cite{openai_codex_review}, an industrial-strength, LLM-based code review tool. We select it as a baseline because it is (i) widely used, (ii) open-source, and (iii) developed by OpenAI, the same organization behind GPT-5-mini, our underlying LLM.
Similar to Testora, we also configure Codex CLI to use GPT-5-mini and apply its \code{/review} feature to analyze the subject PRs.

\paragraph{Developer-written regression tests (\rqone and \rqtwo)}
Developers commonly rely on regression tests to ensure that software evolution does not break existing functionality.
Thus, to assess the quality of the patch oracles generated by \patchguru, we compare them against the developer-written regression tests associated with each project.
Due to SciPy's recent migration from \code{dev.py} to \code{spin}, we are only able to execute regression tests for the 29 most recent PRs.\footnote{The migration caused misconfigured default test commands and test failures,  requiring costly version-specific customization to address.}
To ensure an efficient comparison, we run only those tests that execute the function modified in the analyzed patch.

\subsection{\rqone: Effectiveness at Finding Real-World Bugs}
\label{sec:rq1}

We apply \patchguru to the \datasize PRs and manually analyze raised warnings to identify true inconsistencies between code patches and developer intent expressed in NL artifacts.
To ensure accuracy, warnings were reviewed by the first author and cross-validated by two other authors; any disagreements are resolved through discussion.
A true positive is a warning that correctly identifies such an inconsistency; any other warning is a false positive.
During manual inspection, we find that some true positives exhibit desirable behavior despite being inconsistent with stated developer intent (e.g., inadvertently fixing an unrelated bug). We therefore classify true positives into: (i)~\textit{documentation bugs}, where documentation inaccurately describes the patch's unintended but desirable behavior, and (ii)~\textit{code bugs}, where the patch introduces unintended and undesirable behavior.
We report code bugs that still exist in the latest version, but exclude documentation bugs because all dataset PRs are already merged and developers are unlikely to revise the PR descriptions after merging.

\textbf{Overall Results.}
Table~\ref{tab:bug-detection} summarizes the effectiveness of \patchguru and \testora in detecting real-world bugs across the four studied projects.
Out of \datasize PRs from four projects, \patchguru successfully infers patch oracles for 336 PRs, yielding a success rate of 84\%. 
The success rate is also quite consistent across different projects, ranging from 80\% (Keras) to 90\% (SciPy). 
Unsuccessful inferences are due to
(1)~unresolvable runtime errors (33/64 cases) and
(2)~unresolvable LLM query errors (31/64), mostly caused by token limits when generating comparison programs and lack of adherence to the expected format.
This limitation is expected given current LLM capabilities and the difficulty of generating executable comparison programs for arbitrary real-world patches~\cite{Lars2025ChangeGuard}. We therefore restrict the remaining analysis to the 336 successfully inferred oracles. 
Using these oracles, \patchguru raises \numofwarnings warnings across the four projects, of which \numoftp are true positives and \numoffp are false positives, yielding a \precision precision.

\begin{table}[t]
\centering
\caption{Effectiveness of \patchguru and \testora in detecting real-world bugs. \#PRs, \#Ora., \#Warn, \#TP, and Prec.\ denote the number of analyzed PRs, generated oracles, issued warnings, true positives, and precision.}
\label{tab:bug-detection}
\setlength{\tabcolsep}{4pt}
\resizebox{\columnwidth}{!}{
    \begin{tabular}{@{}lrrrrrrrr@{}}
    \toprule
    \textbf{Project} & \textbf{\#PRs} & \multicolumn{4}{c}{\textbf{\patchguru}} & \multicolumn{3}{c}{\textbf{\testora}} \\
    \cmidrule(lr){3-6} \cmidrule(lr){7-9}
    & & \textbf{\#Ora.} & \textbf{\#Warn} & \textbf{\#TP} & \textbf{Prec.} & \textbf{\#Warn} & \textbf{\#TP} & \textbf{Prec.} \\ \midrule
    Keras       & 100 & 80  & 9  & 5 & 0.56 & 0  & 0 & N/A  \\
    Marshmallow & 100 & 83  & 7  & 4 & 0.57 & 8  & 2 & 0.25 \\
    Pandas      & 100 & 83  & 11 & 8 & 0.73 & 10 & 3 & 0.30 \\
    SciPy       & 100 & 90  & 12 & 7 & 0.58 & 4  & 2 & 0.50 \\ \midrule
    \textbf{Overall} & \textbf{400} & \textbf{336} & \textbf{39} & \textbf{24} & \textbf{0.62} & \textbf{22} & \textbf{7} & \textbf{0.32} \\ \bottomrule
    \end{tabular}
}

\end{table}

\textbf{True Positives.}
Table~\ref{tab:bugs-summary} summarizes the 24 true positives identified by \patchguru.
Among these, \numofcodebug are code bugs and \numofdocbug are documentation bugs. 
Notably, of the \numofcodebug code bugs, \numofunkbug were still present in the latest version of these projects when \patchguru discovered them, while the remaining \numofknownbug had been fixed independently by developers (4) or belonged to functions that were subsequently deleted in later versions (1).
We reported the \numofunkbug previously unknown bugs to the respective project maintainers; as of this writing, maintainers confirmed all these bugs and fixed \numoffix of them.
The only one remaining unfixed bug is in the SciPy project, in which maintainers acknowledged the issue and proposed a fix, but the fix has not yet been implemented. It is worth noting that all PRs analyzed by \patchguru had already been merged into the main codebase, i.e., these issues were not detected by standard patch validation processes, including both regression testing and manual code review. This demonstrates \patchguru's potential to improve software quality assurance by detecting real-world bugs missed by existing processes.

\begin{table}[t]
\centering
\caption{Real-world bugs detected by \patchguru. PR refers to the pull request under analysis.
}
\label{tab:bugs-summary}
\setlength{\tabcolsep}{3pt}
{%
\begin{tabular}{@{}rllll@{}}
\toprule
\textbf{ID} & \textbf{Repo} & \textbf{PR} & \textbf{Kind} & \textbf{Status} \\
\midrule
1 & Keras & 20626 & Code & Confirmed and fixed\\
2 & Keras & 20765 & Code & Confirmed and fixed\\
3 & Keras & 20879 & Documentation & -- \\
4 & Keras & 20928 & Code & Confirmed and fixed\\
5 & Keras & 20974 & Code & Confirmed and fixed\\
\midrule
6 & Marshmallow & 1399 & Documentation & -- \\
7 & Marshmallow & 2698 & Code & Fixed independently \\
8 & Marshmallow & 2699 & Code & Not in latest version \\
9 & Marshmallow & 2800 & Code & Confirmed and fixed\\
\midrule
10 & Pandas & 60828 & Code & Fixed independently \\
11 & Pandas & 61054 & Code & Confirmed and fixed \\
12 & Pandas & 61162 & Documentation & -- \\
13 & Pandas & 61183 & Documentation & -- \\
14 & Pandas & 61646 & Code & Confirmed and fixed\\
15 & Pandas & 61946 & Code & Fixed independently \\
16 & Pandas & 61966 & Code & Confirmed and fixed\\
17 & Pandas & 62085 & Code & Confirmed and fixed\\
\midrule
18 & SciPy & 22213 & Code & Confirmed and fixed\\
19 & SciPy & 22475 & Code & Fixed independently \\
20 & SciPy & 22532 & Documentation & -- \\
21 & SciPy & 22989 & Documentation & -- \\
22 & SciPy & 23280 & Documentation & -- \\
23 & SciPy & 23341 & Code & Confirmed and fixed\\
24 & SciPy & 23520 & Code & Confirmed \\
\bottomrule
\end{tabular}

}
\end{table}

We find that \patchguru effectively identifies three bug types. 
First, it detects \textit{incomplete patches} that fail to fully implement the intended behavior. 
In Pandas PR~\#61966, \patchguru inferred the oracle \code{pre\_out == post\_out[1:-1]} and generated inputs showing that the patch correctly handled \code{str} categories but not \code{string} categories, a bug later confirmed by Pandas maintainers. 
This case also highlights the benefit of patch oracles on cross-version properties.
Second, \patchguru detects \textit{regressions}. 
In Keras PR~\#20765, a fix for partial batch sizes in \code{TimeDistributed} caused the post-patch version to silently accept masks with mismatched timesteps; \patchguru inferred that both versions should raise an error and exposed the regression.
Finally, \patchguru detects \textit{documentation bugs} where NL artifacts inaccurately describe patch behavior.
In Pandas PR~\#61162, the description stated that the patch enables \code{Series[bool]} indexing without specifying that it applies only to integer-indexed series.
\patchguru generated oracles covering both integer- and non-integer indexes, revealing that the post-patch version raises exceptions for the latter, a discrepancy between documented and actual behavior.

\textbf{False Positives.} To understand the current limitations of \patchguru, we manually analyze all \numoffp false positives identified during our evaluation. 
This analysis reveals two primary root causes. 
First, in 8 cases, \patchguru generates unrealistic inputs in the synthesized patch oracles that do not reflect actual usage scenarios, thereby producing false warnings. 
This issue is caused by limitations in \patchguru's current comparison program synthesis and input generation strategies.
Specifically, \patchguru occasionally synthesizes unrealistic mock data types 
to enable the execution of comparison programs, or generates inputs that violate preconditions along the calling stack from top-level APIs to the modified functions. 
Second, in three cases, \patchguru generates incorrect assertions, which misinterpret the developer intent. 
Notably, two cases arise because \patchguru disagrees with developer intent as expressed in NL artifacts despite correctly inferring it.
For example, in PR~\#60867, \patchguru infers that the intended behavior is to return \code{frozenset} objects with curly braces even for empty sets, but assumes that an empty \code{frozenset} should be represented as \code{frozenset()} as opposed to \code{frozenset(\{\})}, resulting in false warnings. 
In two additional cases, \patchguru fails to capture environment-specific behavior, leading to incorrect oracles.
For the remaining two cases, insufficient context prevents us from confirming true positives, so we conservatively classify them as false positives.

\textbf{Comparison with \testora.} 
As shown in Table~\ref{tab:bug-detection}, \testora issued 22 warnings, of which 7 were true positives, yielding a precision of 0.32. 
In contrast, \patchguru detected substantially more true bugs (\numoftp{} vs.\ 7), achieved higher precision (0.62 vs.\ 0.32), and found 22 bugs missed by \testora.
Moreover, 22 of the \numoftp bugs detected by \patchguru are unique to \patchguru and were not identified by \testora.
This performance improvement arises from two factors. 
First, \patchguru infers explicit patch oracles that check both unintended behavioral changes and intended behavior described in NL artifacts, whereas \testora relies on differential testing between pre- and post-patch versions.
For the example in Fig.~\ref{fig:example}, both versions exhibit the same incorrect behavior for uppercase file URL schemes, causing \testora to miss the bug, while \patchguru's inferred patch oracles capture the intended changes and detect the issue. 
Second, \patchguru refines oracles using execution feedback and self-review, improving their alignment with developer intent. 
By contrast, \testora relies on LLM-based interpretation of behavioral differences, which is more prone to hallucination.
This advantage is particularly evident by the higher precision of \patchguru over \testora (0.62 vs.\ 0.32).
Despite these advantages, \testora still detects 5 bugs missed by \patchguru. 
These bugs are missed because \patchguru either fails to resolve runtime errors or generate incomplete patch oracles.
These results suggest that while \patchguru significantly improves bug detection compared to \testora, there are still opportunities for further enhancing its capabilities.

\begin{table}[t]
\centering
\caption{Effectiveness of Codex Code Review on the \numoftp{} buggy PRs detected by \patchguru and on 40 randomly sampled PRs (10 per project). \#Warn and \#TP denote issued warnings and true positives, i.e., actual bugs, respectively.}
\label{tab:codex-comparison}
\setlength{\tabcolsep}{4pt}
\resizebox{0.9\columnwidth}{!}{%
\begin{tabular}{@{}lrrrrrrr@{}}
\toprule
\textbf{Project}
  & \multicolumn{3}{c}{\textbf{\numoftp{} buggy PRs}}
  & \multicolumn{3}{c}{\textbf{40 sampled PRs}} \\
\cmidrule(lr){2-4}\cmidrule(lr){5-7}
  & \textbf{\#Warn} & \textbf{\#TP} & \textbf{Precision}
  & \textbf{\#Warn} & \textbf{\#TP} & \textbf{Precision} \\
\midrule
Keras       & 5 & 1 & 0.2 & 7 & 0 & 0.0 \\
Marshmallow & 4 & 4 & 1.0 & 2 & 0 & 0.0 \\
Pandas      & 8 & 4 & 0.50 & 7 & 2 & 0.29 \\
SciPy       & 6 & 3 & 0.50 & 7 & 0 & 0.0 \\
\midrule
\textbf{Overall} & 23 & 12 & 0.52 & 23 & 2 & 0.09 \\
\bottomrule
\end{tabular}%

}
\end{table}

\textbf{Comparison with Codex CLI.}
Finally, we compare \patchguru against the \texttt{/review} feature of Codex CLI~\cite{openai_codex_review}; the results are summarized in Table~\ref{tab:codex-comparison}.
Since it produces many warnings per PR, manually labeling all warnings across all \datasize{} PRs is infeasible. 
We therefore evaluate it in two parts: (i)~we run it on the \numoftp{} PRs where \patchguru detected a bug to see how well \patchguru complements Codex; and 
(ii)~we randomly sample 10 PRs per project from the remaining PRs and inspect all warnings to estimate Codex's precision.

\emph{Complementarity of \patchguru to Codex Code Review.}
On the \numoftp{} PRs where \patchguru detected a bug, Codex raised 23 warnings with 12 true positives (precision 0.52), missing 12 of the \numoftp{} bugs found by \patchguru.
This demonstrates that \patchguru's patch-oracle-based approach catches bugs that Codex misses, showing the potential for \patchguru to complement existing LLM-based code review tools.

\emph{Precision of Codex Code Review.}
On 40 randomly sampled PRs, Codex raised 23 warnings but only 2 true positives, yielding a precision of 0.09 and requiring substantial manual triage.
We attribute this advantage to \patchguru's use of patch oracles, which check whether a patch satisfies its intended behavior, whereas Codex Code Review relies on LLM-based reasoning that remains prone to false positives due to limited deep code-semantic understanding~\cite{LeCong2025FormalBench, lin2025codereviewqa, liu2026assessing}.
Overall, these results indicate that \patchguru can complement existing LLM-based code review tools by improving bug detection and producing more actionable warnings.

\subsection{\rqtwo: Adequacy of Patch Oracles}

Besides detecting real-world bugs, another crucial aspect of patch oracles is their adequacy, which directly impacts their effectiveness in validating code patches. 
In \rqtwo, we evaluate the adequacy of patch oracles generated by \patchguru using mutation testing~\cite{jia2010analysis,papadakis2019mutation}, a well-established approach for assessing test quality.
Specifically, we leverage Mutmut~\cite{hovmoller2016mutmut}, a mutation testing framework for Python, to generate mutants with all default mutation operators.
Mutants are generated on the \emph{post-patch} version to simulate patch-induced bugs. 
Test adequacy is measured by the mutation score, i.e., proportion of mutants killed, with higher scores indicating better adequacy.

\begin{figure}[t]
 \centering
 \includegraphics[width=0.9\columnwidth]{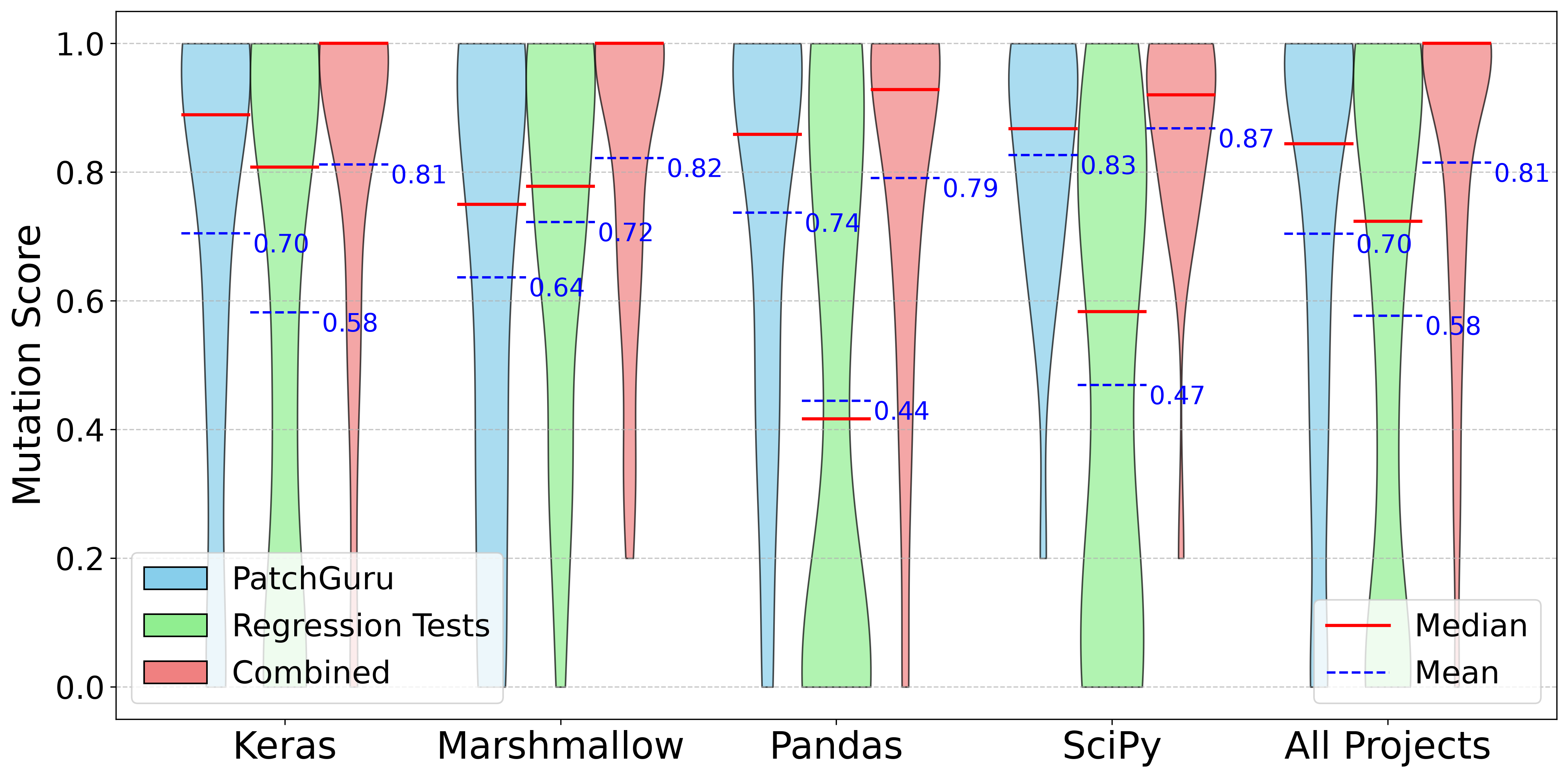}
 \caption{Distribution of mutation scores for \patchguru-generated oracles and developer-written regression tests.}
 \label{fig:spec-quality}
 
\end{figure}

\begin{figure*}[t]
\centering
\begin{subfigure}[t]{0.32\textwidth} %
    \centering
    \includegraphics[width=\linewidth]{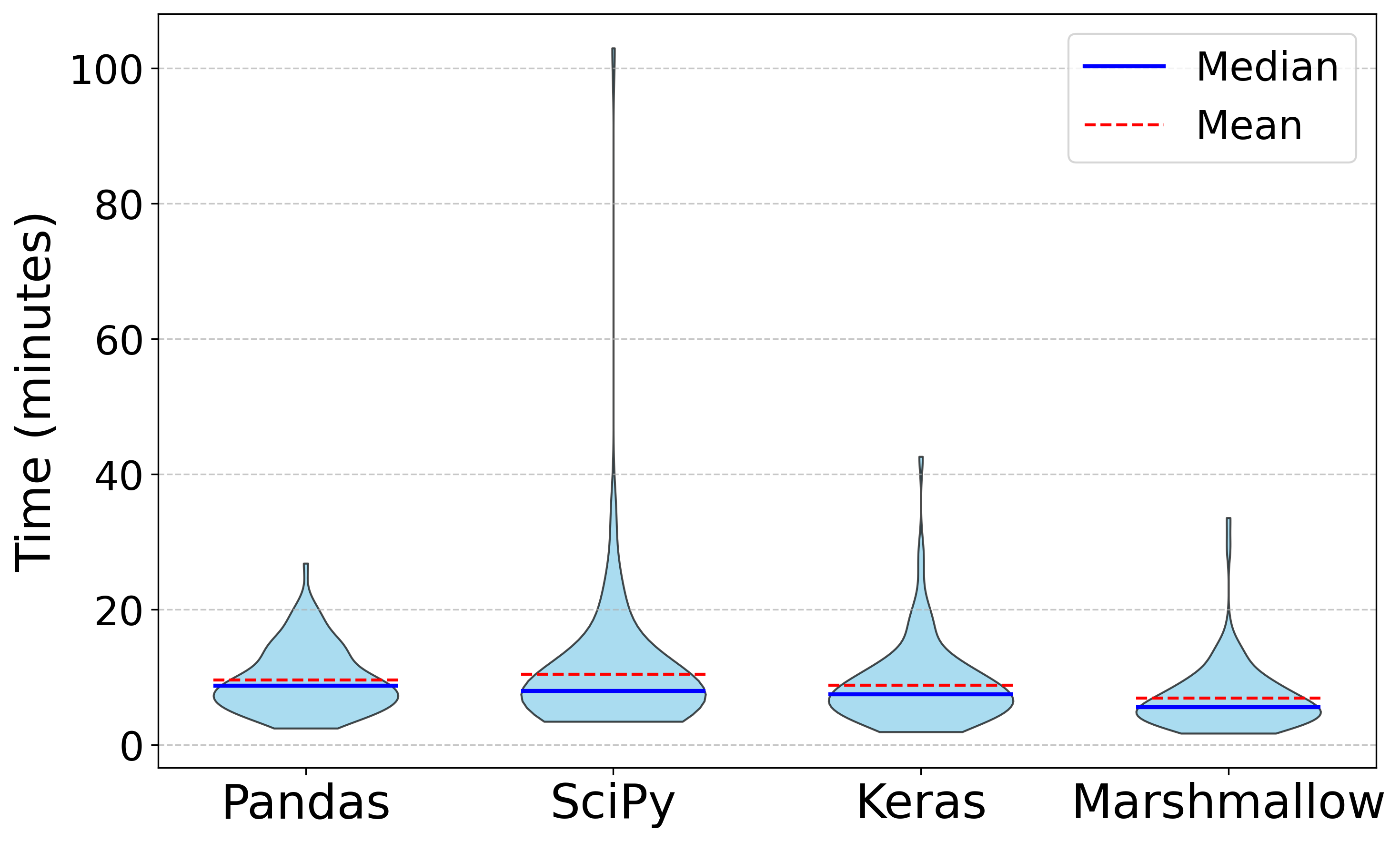}
    \caption{Time}
    \label{fig:execution-time-distribution}
\end{subfigure}
\hfill
\begin{subfigure}[t]{0.32\textwidth}
    \centering
    \includegraphics[width=\linewidth]{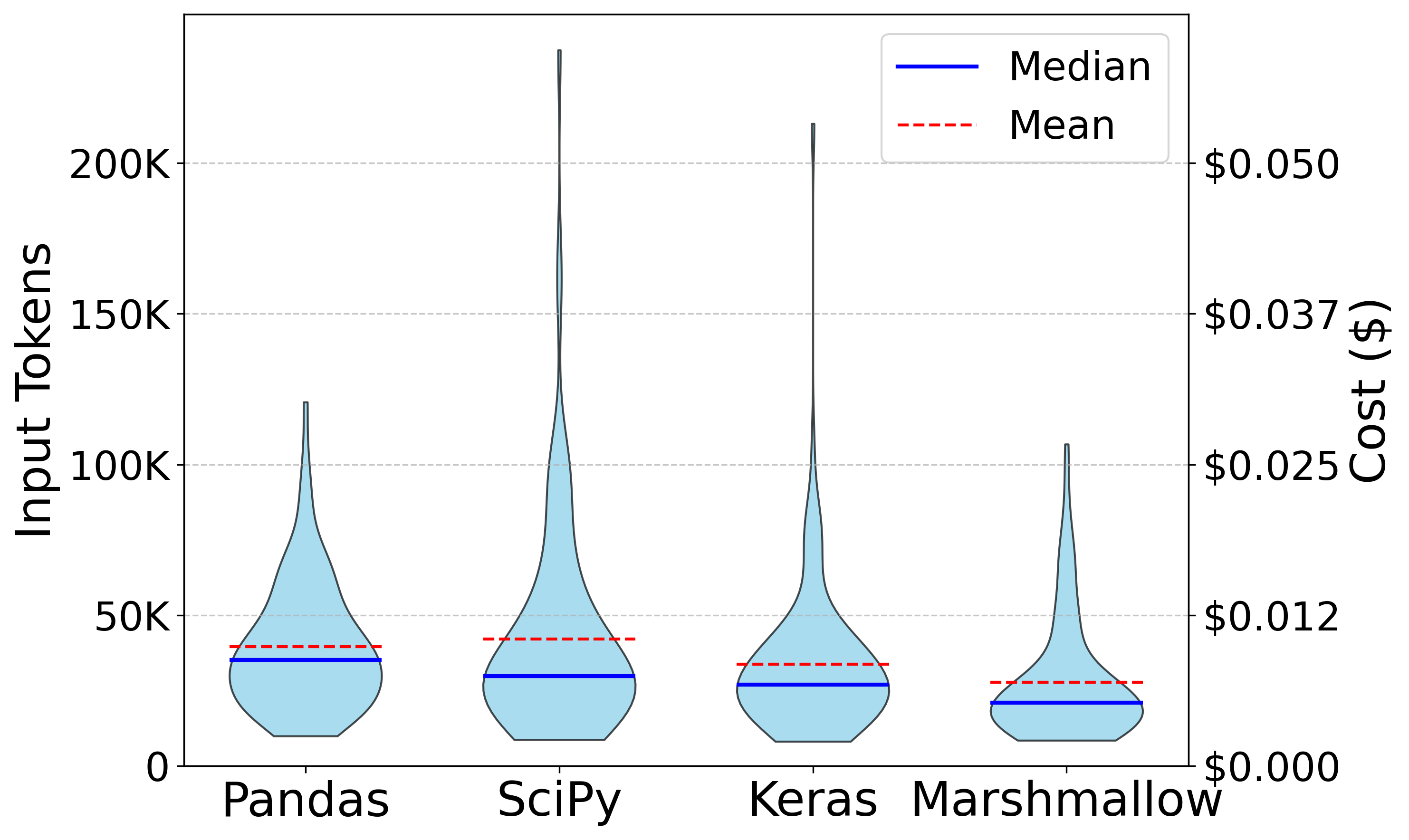}
    \caption{LLM input tokens}
    \label{fig:input-token-usage-distribution}
\end{subfigure}
\hfill
\begin{subfigure}[t]{0.32\textwidth}
    \centering
    \includegraphics[width=\linewidth]{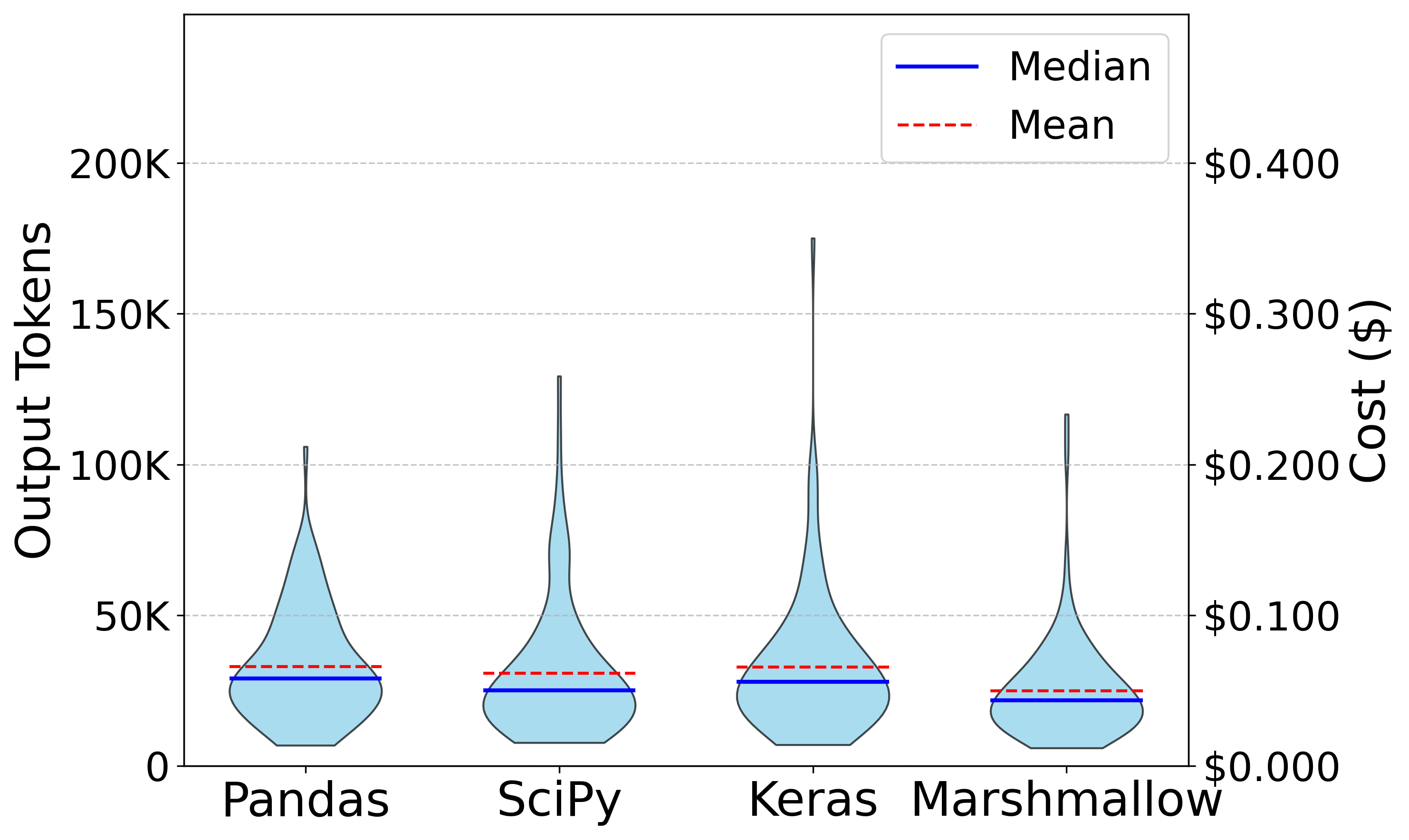}
    \caption{LLM output tokens}
    \label{fig:output-token-usage-distribution}
\end{subfigure}
\caption{Execution time and token usage per PR.}
\label{fig:efficiency-metrics}
\end{figure*}

\textbf{Results.}
Fig.~\ref{fig:spec-quality} compares the distribution of mutation scores achieved by \patchguru-generated oracles, developer-written regression tests and their combination across the four studied projects.
Overall, generated patch oracles are more adequate than developer-written regression tests, achieving 21\% higher overall mutation scores (0.70 vs. 0.58). 
A Wilcoxon signed-rank test confirms that the improvements are statistically significant for Keras, Pandas, and SciPy (p-value $<$ 0.05). 
Moreover, combining \patchguru-generated oracles with developer-written regression tests substantially improves the mutation score, to an overall value of 0.81.
Median scores of the combined oracles even reach 1.0 in Pandas and Marshmallow. 
These results show that \patchguru can effectively complement existing regression tests.

\subsection{\rqthree: Efficiency}

In \rqthree, we evaluate the computational cost of \patchguru using two metrics: the analysis time per PR and the token usage of LLM calls.
Because LLM costs vary over time and across providers, we measure input and output token counts, which directly correlate with incurred costs.

\textbf{Execution Time.}
On average per checked PR, our approach takes 8.9 minutes to complete the entire analysis. 
Fig.~\ref{fig:execution-time-distribution} illustrates the distribution of execution times across all analyzed PRs. 
We can see that most PRs are analyzed within 20 minutes, with a few outliers taking longer due to the complexity of the patches or the need for multiple iterations of oracle enhancement. 
We can also see that the distribution of execution times is relatively consistent across different projects with average times ranging from 6.9 minutes (Marshmallow) to 10.4 minutes (Pandas). 
These results suggest that \patchguru can analyze real-world PRs efficiently and within a reasonable time frame, making it feasible for practical CI/CD deployment.

\textbf{Monetary Costs.} On average per checked PR, \patchguru utilizes 35,666 input tokens and 30,230 output tokens, equivalent to approximately USD~0.009 and USD~0.06, respectively, in LLM calls. 
Figures~\ref{fig:input-token-usage-distribution} and \ref{fig:output-token-usage-distribution} illustrate the distribution of token usage across all analyzed PRs. We can see that most PRs require fewer than 60K input tokens and 50K output tokens. 
As for execution times, the token usage distribution is relatively consistent across different projects, with average input tokens ranging from 27,644 (Marshmallow) to 42,115 (SciPy) and average output tokens ranging from 24,846 (Marshmallow) to 32,694 (Keras). 
Based on the pricing of OpenAI's GPT-5-mini model as of June 2026,
the total cost per PR is estimated to be approximately USD~0.07. 
We believe this cost is acceptable for practical use, especially considering the benefits of early bug detection and improved patch validation.

\subsection{\rqfour: Ablation Study}

\rqfour conducts an ablation study to assess the contribution of individual components of \patchguru to its overall performance. Specifically, we evaluate two modules: (i) the oracle enhancement module (\S\ref{sec:approach_enhancement}) and (ii) the self-review module (\S\ref{sec:approach_review}).
We create two ablation variants of \patchguru by removing each component in turn and evaluating them on the same set of \datasize PRs.
The results are reported in Table~\ref{tab:ablation-study}.

\textbf{\patchguru without Oracle Enhancement.} We first disable the oracle enhancement module and directly use the initial patch oracles generated by the LLM without further refinement. 
As shown in Table~\ref{tab:ablation-study}, removing this module significantly reduces \patchguru's effectiveness in detecting real-world bugs: the number of warnings drops from \numofwarnings to \numofwarningspone, and true positives decrease from \numoftp to \numoftppone.
Therefore, the oracle adequacy also decreases, with the mutation score dropping from \mutationscore to \mutationscorepone. 
Although precision slightly increases from \precision to \precisionpone, the bug detection is considerably affected.
These results demonstrate that the enhancement module is essential for improving oracle quality by generalizing patch oracles and generating comprehensive test inputs.

\begin{table}[t]
\centering
\caption{Ablation study results. * indicates estimated values.}\label{tab:ablation-study}
\resizebox{\columnwidth}{!}{
    \begin{tabular}{@{}lcccc@{}}
    \toprule
    \textbf{Method} & \textbf{\#Warnings} &\textbf{\#Bugs} & \textbf{Precision} & \textbf{Mutation Score} \\ \midrule
    \patchguru & \numofwarnings & \numoftp & \precision & \mutationscore \\ 
    w/o oracle enhancement & \numofwarningspone & \numoftppone & \precisionpone & \mutationscorepone \\
    w/o self-review &  195 & \;25* & \;0.13* & N/A \\ \bottomrule
    \end{tabular}
}

\end{table}

\textbf{\patchguru without Oracle Self-Review.} 
Then, we disable the self-review module and directly report all detected inconsistencies, yielding 195 warnings across 400 PRs, making exhaustive manual inspection impractical.
We therefore estimate results by partitioning warnings based on the full \patchguru's self-review outcomes: for warnings classified as ``bug'', we reuse the inspection results from \rqone; for warnings filtered out, we randomly sample five per project, manually inspect them, and extrapolate to the full set.
Based on this estimation, we found that disabling the self-review module causes a substantial increase in false positives, leading to a sharp drop in precision from \precision to 0.13.
Of the 20 sampled warnings filtered by the self-review module, 19 were confirmed as false positives upon manual inspection with common false-positive patterns including incorrect comparison operators, invalid assumptions about pre-patch behavior, incorrect invariants, and improper input initialization.
Although the estimated number of detected bugs increases slightly from \numoftp to 25, this gain comes at the cost of an overwhelming number of false positives, rendering the tool impractical for developers.

\section{Threats to Validity}
\label{sec:discussion}

\textbf{Internal Validity.}
First, our manual classification into true and false positives is subject to human error; we mitigate this threat through cross-validation among authors and independent confirmation from project maintainers.
Second, \patchguru assumes the pre-patch function is correct; while this may cause false negatives when the pre-patch itself is buggy, it is consistent with \patchguru's goal of detecting bugs \emph{introduced} by the patch.
Finally, LLM non-determinism threatens reproducibility~\cite{Sallou2024LLMNondeterminism}. However, we argue that this is inherent in LLMs and reflects the variability that practitioners encounter in real-world usage. 
We minimize this risk by conducting a large-scale evaluation on 400 PRs across four projects and providing all intermediate artifacts in the replication package~\cite{Anom2025PatchGuruReplicationPackage}.

\textbf{External Validity.}
Our benchmarks are not representative of all software projects.
We mitigate this by selecting popular Python projects from diverse domains following Testora~\cite{Pradel2025TestoraUN}. 
However, \patchguru currently targets Python projects,  which may limit the generalizability of our findings to other languages.
Although its core approach is language-agnostic, supporting additional languages may require substantial engineering effort.
Finally, \patchguru currently handles only single-function patches, which simplifies comparison program construction.
Extending it to multi-function patches is non-trivial because it requires accurately modeling interactions among modified functions and their dependencies.
We acknowledge this as a limitation and leave it as future work.

\textbf{Construct Validity.}
We use mutation score as a proxy for oracle adequacy.
While mutation testing is a widely accepted technique for assessing test suite quality~\cite{jia2010analysis,papadakis2019mutation}, mutation score may not perfectly correlate with real-world fault detection capability. 
To mitigate this threat, we complement mutation score with evaluation on real-world PRs, which provides a more direct assessment of \patchguru's effectiveness in practice.
The effectiveness of \patchguru also depends on LLM capabilities; due to limited funding, we evaluated only GPT-5-mini.
Future work could assess performance across diverse LLMs and configurations.

\section{Related Work}
\label{sec:related_work}

\textbf{Reasoning about Code Changes.}
As software evolves, reasoning about code changes is crucial for maintaining software quality.
The most relevant line of work to \patchguru is on patch specification~\cite{Cadar2023PatchSV,Sharma2025P3Reasoning} and software change contracts~\cite{SoftwareChangeContracts2015}, 
which capture the intended behavior of code changes but require manually provided specifications.
\patchguru addresses this by automatically inferring patch oracles, 
an under-approximate yet practical form of patch specifications, from NL artifacts, enabling automated validation with minimal human intervention.
\patchguru is also closely related to recent work on reasoning about code changes using learning-guided execution and LLMs~\cite{Lars2025ChangeGuard,Pradel2025TestoraUN}. 
Compared to ChangeGuard~\cite{Lars2025ChangeGuard}, which introduced the idea of comparison programs, we go beyond determining whether a code change introduces any behavioral differences but instead reason about the consistency of the intent expressed in the NL artifacts and the code change.
Compared to Testora~\cite{Pradel2025TestoraUN}, our formulation not only checks whether PRs that introduce unintended behavioral changes, but also validates that the intended changes are correctly implemented as demonstrated in \S\ref{sec:evaluation}. 
\patchguru is also related to just-in-time bug prediction~\cite{hoang2019deepjit, zeng2021deep, nguyen2025toward, nguyen2025vulguard, perl2015vccfinder} which estimates defect likelihood from historical data. However, recent works~\cite{nguyen2025toward, zeng2021deep, fan2019impact} show that these approaches have limited practical effectiveness due to class imbalance and noisy SZZ labels.
More broadly, automated testing techniques have been used to target patches~\cite{kuchta2018shadow,aflgo,dafl,directed-symex11, icse2026-ChaCo}. 
While these techniques effectively generate tests that exercise code changes, they do not address the oracle problem, which is essential for determining correctness.
In principle, they are complementary to \patchguru and could be combined with it to generate tests that are then validated against inferred patch oracles.

\textbf{Test Oracle Problem.}
Software testing relies on test oracles to determine the correctness of program outputs, yet such oracles are often incomplete or absent~\cite{barr2014oracle}.
Prior work addresses this challenge by inferring oracles from API documentation, including metamorphic relations~\cite{blasi2021memo} and exceptional behavior~\cite{goffi2016automatic};
or
using deep learning models and LLMs~\cite{dinella2022toga,hossain2025doc2oracll}.
Other approaches infer program specifications that serve as oracles through dynamic invariant inference~\cite{nguyen2012using,ernst2007daikon,le2023invalidator} or LLM-based techniques~\cite{LeCong2025FormalBench,endres2024can}.
Despite their effectiveness, these techniques primarily validate overall program behavior rather than behavior introduced or modified by code changes. 
This formulation is ill-suited to large, rapidly evolving systems because it requires comprehensive behavioral coverage, which is often infeasible in practice, and continual oracle updates, which impose substantial maintenance costs.
In contrast, \patchguru infers patch oracles that target only the behavior affected by a change, enabling more efficient validation.

\section{Conclusion}
\label{sec:conclusion}

Software patches are essential for system evolution but remain a major source of bugs and vulnerabilities due to a lack of machine-checkable specifications capturing developer intent.
In this paper, we introduced \patchguru, an automated technique that infers machine-checkable patch oracles from natural language artifacts using LLMs and dynamic analysis.
Our evaluation on \datasize real-world PRs shows that \patchguru identifies \numoftp real inconsistencies, including \numofunkbug previously unknown code bugs.
The approach imposes reasonable costs of 8.9 minutes and \$0.07 per PR, suitable for real-world deployment.
We envision \patchguru as a complement to existing development practices by providing machine-checkable behavioral documentation and enabling early detection of unexpected behaviors missed by traditional approaches.

\bibliographystyle{IEEEtran}
\bibliography{main}
\balance

\end{document}